\documentclass[twocolumn,showpacs,preprintnumbers,pre]{revtex4}
	\usepackage[utf8]{inputenc}  
	\usepackage[iso]{datetime}
	\usepackage{amsmath, amssymb}
	\usepackage{xcolor}
	\usepackage{graphicx,epstopdf}
	\usepackage{multirow}
	\usepackage[caption=false]{subfig}
	  
	\newcommand{\reffig}[1]{Fig.~\ref{#1}}
	
	\newcommand{\reftbl}[1]{Table~\ref{#1}}

\usepackage{textcomp}
	\usepackage{amsmath, amssymb,amsfonts,amsthm}
		 
	\newcommand{\hbAbs}[1]{| #1 |}

	\newcommand{\argmax}[2]{\underset{#1}{\operatorname{arg \, max}}\;#2}
	\theoremstyle{plain}

	\theoremstyle{definition}

	\theoremstyle{remark}

	\usepackage{changes}
	\definechangesauthor[name=Haluk Bingol,color=purple]{hb}
	\definechangesauthor[name=Mursel Tasgin,color=blue]{mt}
	\setremarkmarkup{(#2)}
	
\usepackage{clrscode3e} 
\usepackage{float}
\usepackage{array}
\usepackage{booktabs}
\usepackage{xspace} 
\usepackage{textgreek}
\newcommand{\hbAlgorithmCN}{PCN}
\newcommand{\hbAlgorithmSC}{PSC}
		 
	\newcommand{\kMax}{k_{\mathrm{max}}}

\begin{document} 

\title{
	Community detection using preference networks
}
\author{Mursel Tasgin}
\author{Haluk O. Bingol}
\affiliation{Department of Computer Engineering\\
Bogazici University, Istanbul \\
}

\begin{abstract}
Community detection is the task of identifying clusters or groups of nodes in a network
where nodes within the same group are more connected with each other than with nodes in different groups.  
It has practical uses in identifying similar
functions or roles of nodes in many biological, social and computer networks.
With the availability of very large networks in recent years, 
performance and scalability of community detection algorithms become crucial,
i.e. if time complexity of an algorithm is high, it can not run on large networks.
In this paper, we propose a new community detection algorithm, 
which has a local approach and is able to run on large networks.
It has a simple and effective method;
given a network, algorithm constructs a \emph{preference network} of nodes
where each node has a single outgoing edge showing its preferred node to be in the same community with.
In such a preference network, each connected component is a community.
Selection of the preferred node is performed using similarity based metrics of nodes.
We use two alternatives for this purpose 
which can be calculated in 1-neighborhood of nodes,
i.e.
number of common neighbors of selector node and its neighbors and,
the \emph{spread capability} of neighbors around 
the selector node which is calculated by the gossip algorithm of Lind et.al.
Our algorithm is tested on both computer generated LFR networks and real-life networks with ground-truth community structure.
It can identify communities accurately in a fast way.
It is local, scalable and suitable for distributed execution on large networks.
\end{abstract}


\maketitle

\section{Introduction}

Community detection is one of the key areas in complex networks 
that has attracted great attention in the last decade.
In network science, a network is seen as a system and individual nodes as agents 
or elements of the system where they are connected with ties~\cite{wasserman1994social}.
Mobile communication networks, scientific collaboration networks, 
patent networks, protein interaction networks and brain networks are examples of 
network representation of corresponding systems~\cite{%
	onnela2007PNAS,%
	newman2001Collaboration,%
	leskovec2005Patent,%
	chen2006Protein,%
	sporns2013Brain%
}.
Interaction of agents in a network can create emergent structures like communities.
A \emph{community} is defined as grouping of nodes in a network 
such that nodes in the same group have more connections 
with each other than with the nodes in the rest of the network~\cite{%
	girvan2002community}.   
There have been many algorithms proposed so far for community detection and
there is a comprehensive survey by Fortunato on community detection~\cite{%
	fortunato2010Survey}.

While many algorithms perform well on small networks with hundreds or thousands of nodes, 
only a few of them can run on very large networks of millions or billions of nodes due to performance and time-complexity issues.
If a community detection algorithm has to deal with the whole network during its execution steps 
or needs to optimize a global value (i.e. network modularity), 
it becomes computationally expensive to run this algorithm on large networks. 
Besides their large sizes, real-life networks also evolve over time,
i.e. structure and size can change while a community
detection algorithm is still running on such a large network.

In recent years local community detection algorithms are proposed to 
overcome the challenges of large networks.
Local algorithms are scalable and are suitable for distributed run,
i.e. they can work on separate parts of the network locally and then merge results for the whole network.
Their local nature makes it possible to identify more granular structures 
which is useful for finding subtle communities; 
especially in networks of loosely connected groups of nodes.

In this paper, we propose a new community detection algorithm which has a local approach.
Our assumption is that, each node in the network selects to be 
a member of a community in order to be in the same group with some \emph{preferred nodes}.
Whether it is the common things they share, common enemies they avoid, 
common features they have or common ones they follow; 
being a member of a community is meaningful only when friends or preferred nodes are together there.
Such communities can be constructed by asking each node 
who they would like to be with; and then grouping them together according to given answers.
How should a node decide on its preferred node:
a popular node, a hub node connecting others or a node with most common friends?
We think that using the similarity of nodes is a good way to make the decision,
i.e. each node should select the neighbor with whom it has the most similarity.

When we are given a network dataset, generally we only have 
the knowledge of nodes and edges,
 i.e. no meta-data describing 
the nodes or common features of nodes (i.e. similarity) may exist.
So, using the connectivity information (edges) among the nodes in network, 
we can get some of the features of individual nodes (i.e. centrality, degree etc.) 
and some similarity measures between nodes (i.e. common neighbors shared by two nodes).
In order to keep local, we should limit our attention to \emph{local metrics},
 i.e. metrics regarding a node and its local neighborhood only, not further.
Alternatively, we can try to use some other methods to make decision on preferred node,
i.e. selection of the neighbor with highest degree, 
 selection of the neighbor having highest clustering coefficient etc.
We will go into the details of metrics for preferred node selection later.\\

Outline of the paper is as follows.
We will give background information about the topics in community detection and methods we use in our algorithm.
We then explain our community detection approach in detail.
We compare our algorithm with other known algorithms on both real-life networks and computer generated networks.
We will finish with the conclusion section.

\section{Background}

\subsection{Local approach for community detection}

Community formation is something local by nature.
But algorithms to detect communities may use global information rather than local information.
For example, 
one community detection algorithm~\cite{%
	girvan2002community}
iterates over the network.
In each iteration, 
it calculates the number of shortest paths passing through each edge and 
removes the edge with the largest number.
Such calculations require information of the entire network.
This global approach, 
which is fine for networks of small sizes, 
is not feasible for very large networks.

In recent years several local community detection algorithms have been proposed~\cite{%
	newman2004fast,
	rosvall2007Infomap,
	raghavan2007LPA,
	blondel2008Louvain,
	gregory2010COPRA,
	lancichinetti2011OSLOM,
	de2014mixing,
	eustace2015community}.
One popular method is the label propagation algorithm~\cite{raghavan2007LPA} which has linear time-complexity
and nodes decide on their communities according to the majority of their neighbors.
Local community detection algorithms generally discover 
communities based on local interactions of nodes 
or local metrics calculated in their 1-neighborhood.
Some algorithms merge nodes into communities based on the optimization of a local metric~\cite{%
	eustace2015community}.
Besides their scalability, these approaches are also suitable for large networks evolving over time.
A local algorithm can handle what it already has (i.e. a portion of the network at a certain time) 
and can continue with what will come later; it does not need the snapshot of the whole network at once.

\subsection{Triangles and communities in networks}

In his book, Simmel~\cite{%
	simmel1950sociology} 
argued that a 
strong social tie could not exist without being part of a triangle in a relation, 
i.e. relation among three people where all know each other.
People who have common friends are more likely to create friendships; they form triangles.
There is a correlation between triangles and communities in social networks; 
there exists many triangles within communities while very few or 
no triangle exists between nodes of different communities~\cite{%
	radicchi2004Clustering}.
Triangle is the smallest cycle of size 3.
There are studies that investigate cycles of size 4 or more~\cite{%
	lind2007_NJP}
but we focus on triangles in this study.

Clustering coefficient (\emph{CC}), is equal to the probability that two nodes 
that are both neighbors of the same third node 
will be neighbors of one another~\cite{%
	newman2001clustering}.
This metric shows the number of existing triangles around a node compared to the all possible triangles.
A high clustering coefficient will mean many triangles and clustering around a node; 
\[
	CC(i)= \frac{\bigtriangleup_{i}}{\wedge_{i}}
\]
where 
$\bigtriangleup_{i}$ is the number of triangles around node $i$ and 
$\wedge_{i}$ is the number of triplets where $i$ is in center.
A \emph{triplet} is formed by three nodes and two edges 
i.e. a triplet $a, i, b$ centered at $i$ has edges $(a, i)$ and $(b, i)$.
%

\newcommand{\myFigFactor}{1}
\begin{figure}
	\centering
	\includegraphics[scale=\myFigFactor]%
		{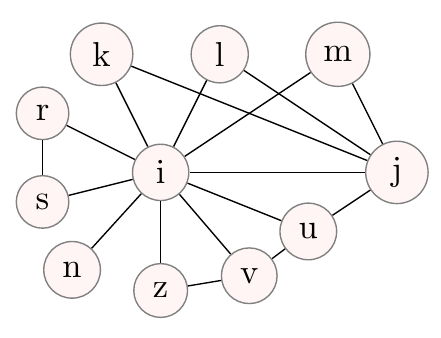}
		
	\caption{
		Clustering coefficient, common neighbors and triangle cascades
	}
	\label{fig:TriangleCascade}
\end{figure} 

Radicchi et.al.~\cite{radicchi2004Clustering}
proposed a community detection algorithm based 
on triangles and clustering coefficient.
In recent years, local community detection algorithms are more focused 
on similarity of nodes; especially on local level, i.e. 1-neighborhood of nodes. 
A node similarity based approach is applied on several known community detection algorithms by 
Xiang et.al.~\cite{xiang2016nodesimilarity},
where they presented the achieved improvements on those algorithms by using various node similarity metrics.
Some examples of those similarity metrics are number of common neighbors and
Jaccard similarity~\cite{jaccard1901Similarity}.

\emph{Number of common neighbors} of nodes $i$ and $j$ is
$CN(i, j) = \hbAbs{\Gamma(i) \cap \Gamma(j)}$,
where $\Gamma(i)$ denotes the
1-neighborhood of $i$.
Number of common neighbors shows the number of triangles formed on two nodes,
i.e. 
in \reffig{fig:TriangleCascade}, 
four triangles are formed on $i$ and $j$ 
by their common neighbors $k$, $\ell$, $m$, and $u$.

\emph{Jaccard similarity} is the fraction of common neighbors of $i$ and $j$ to 
the union of their 1-neighborhoods given as
\[
	J(i, j) = \frac{\hbAbs{\Gamma(i) \cap \Gamma(j)}}
			{\hbAbs{\Gamma(i) \cup \Gamma(j)}}.
\]
All of these metrics are related with friendship transitivity and triangles.

A new metric, spread capability, is proposed as a similarity metric in this paper.
This metric is calculated by using the gossip algorithm of Lind et.al.~\cite{Lind2007PRE}.
A gossip about a \emph{victim} node $i$ is initiated by one of its neighbors, node $j$ (\emph{originator}),
and $j$ spreads the gossip to common friends with $i$,
i.e. gossip about $i$ is meaningful to friends of $i$ only.
Nodes hearing the gossip from $j$ behave the same way 
and propagate it further in the 1-neighborhood of $i$ 
until no further spread is possible.
To measure how effectively the gossip is spread,
they calculate spread factor of victim $i$ by originator $j$ as
\[
	\sigma_{i, j} = \frac
				{\hbAbs{\Gamma_{j}(i)}}
				{\hbAbs{\Gamma(i)}}
\]
where $\Gamma_{j}(i)$ is the set of neighbors of $i$ who heard the gossip originated by $j$.
Lind et.al. calculated the spread factors of each 
originator $j\in\Gamma(i)$ and averaged them to get \emph{spread factor} of $i$, 
i.e.
\[
	\sigma_{i} 
	= \frac{1}{\hbAbs{\Gamma(i)}} 
		\sum_{j \in \Gamma(i)} {\sigma_{i, j}}.
\]

As we will explain in more detail in the next section, 
we use the spread factor in a different way in our algorithm; 
instead of average gossip spread factor of a node, i.e. $\sigma_i$,
we focus on $\sigma_{i, j}$ values, 
which show the contribution of each originator $j$ to that average.
We call $\sigma_{i, j}$ as \emph{spread capability} of $j$ around $i$.
Spread capability is directly related with the connectivity of $j$ and 
its position in the neighborhood of $i$.
So each $j \in \Gamma(i)$ can have a different spread capability around $i$
and
they can be used as a similarity measure between $i$ and $j$ from the perspective of node $i$.
Note that $\sigma_{i,j}\neq\sigma_{j,i}$.

\begin{figure}
	\centering
	\subfloat[\label{fig:CCvsSFtriangleCascades}]{
		\includegraphics
			[scale=.1]
			{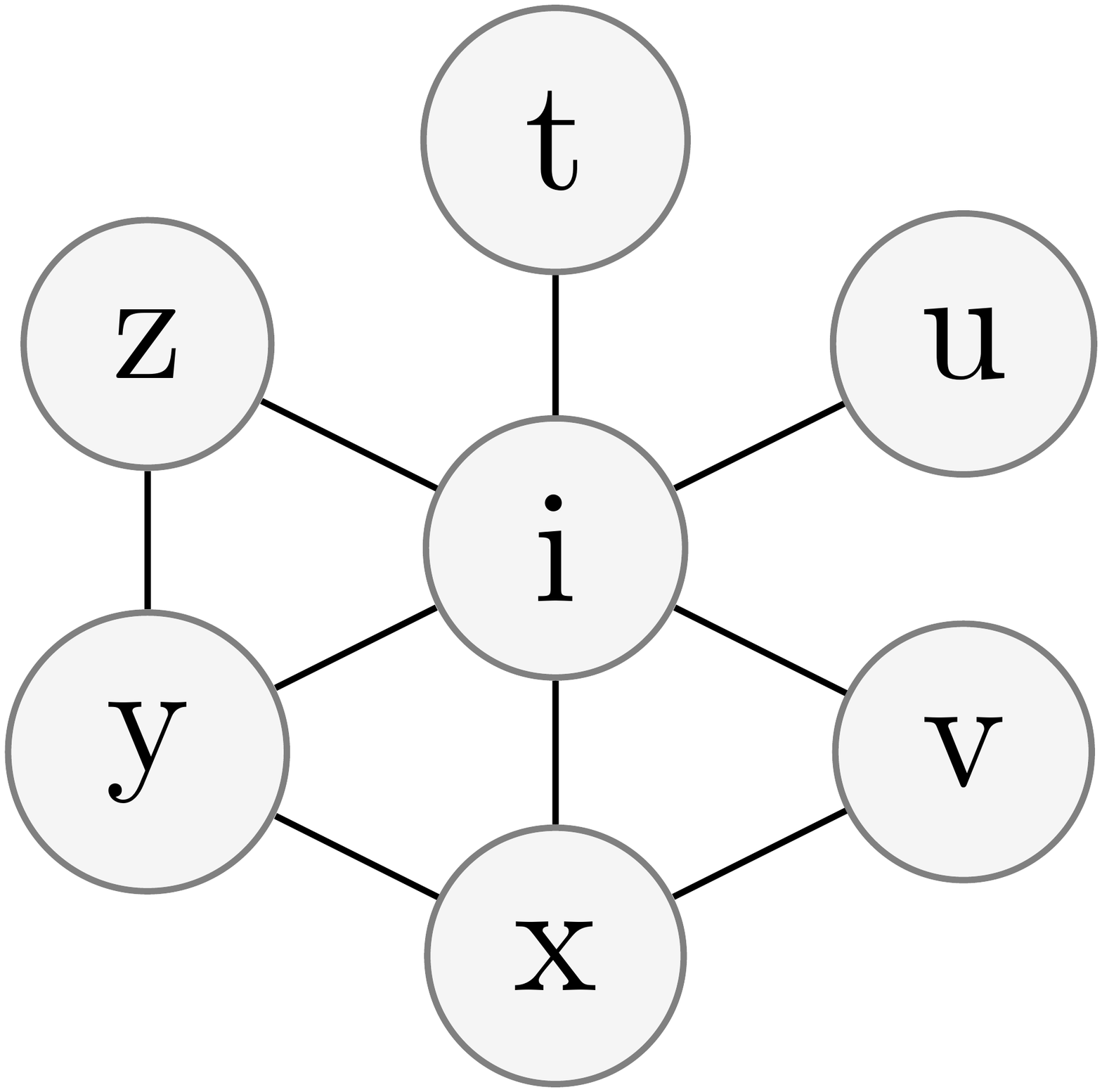}
	}
	\subfloat[\label{fig:CCvsSFclusters}]{
		\includegraphics
			[scale=.1]
			{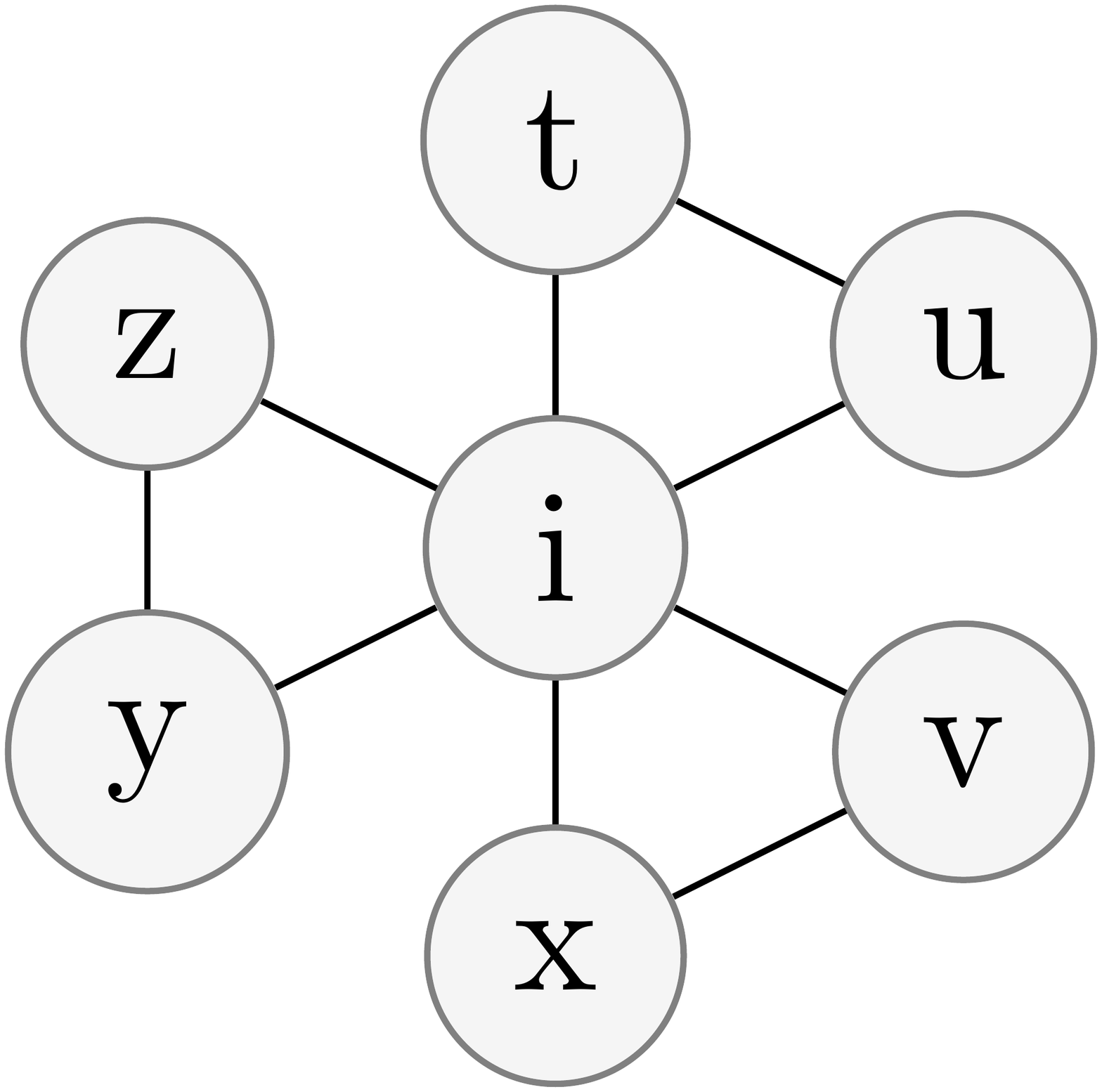}
	}
	\caption{
	Networks (a) and (b) have 7 nodes and 9 edges.
	Degree of $i$ is 6 and clustering coefficient of $i$ is $0.2$ in both cases.
	But average spread factors are 
	$0.5$ in (a) and
	$0.33$ in (b).
	}
	\label{fig:CCvsSFexample}
\end{figure}

Spread capability metric has similarity with number of common neighbors and clustering coefficient; 
but has additional information, see \reffig{fig:CCvsSFexample}.
It contains the number of common neighbors (triangles) between $i$ and $j$; 
moreover
it has the number of other triangles around $i$ with its neighbors
along the spreading pathway of gossip originated by $j$.
We call such adjacent group of triangles as a \emph{triangle cascade},
where all triangles are cornered at the same node (i.e. $i$)
and are adjacent to each other through common edges.

In \reffig{fig:TriangleCascade}, 
$j,u,v,z \in \Gamma(i)$ form a triangle cascade cornered at $i$.
On this triangle cascade, gossip about $i$ originated by $j$ is spread to $k,l,m,u$ directly. 
By using the cascade, gossip is propagated to $v$ by $u$ and then to $z$ by $v$.
Although $v,z \notin \Gamma(j)$,
$j$ still has a role in spreading gossip to $v$ and $z$
by means of triangle cascades.
Hence we have $\Gamma_{j}(i) = \{ k, \ell, m, u, v, z \}$.
Note that for all $a \in \Gamma_{j}(i)$,
we have $\Gamma_{a}(i) = \Gamma_{j}(i)$.
This property becomes very useful to reduce computation of gossip spread factor on a cascade.
Once $\sigma_{i, j}$ is calculated,
then
$\sigma_{i, j} = \sigma_{i, a}$ for all $a \in \Gamma_{j}(i)$.
See further discussion in SI~\cite{Tasgin2017SI}.

\subsection{Method for comparison of two partitions}

Success of a community detection algorithm
lies in finding communities of ground-truth.
Since every node belongs to exactly one community in our algorithm,
being in the same community is an equivalence relation.
Hence, community structure of a network is a partition of a set of nodes.
Suppose we have the partition of the ground-truth.
A community detection algorithm produces another partition.
Then we need a way of comparing these two partitions.
For comparison of two partitions, \emph{Normalized Mutual Information} (NMI)
 can be used~\cite{%
 	danon2005NMI}.
NMI is a metric to understand how far (close) two partitions are; 
if NMI of two partitions is close to 1, then they are very similar,
i.e. number of communities and the members of communities in two partitions are similar;
and when it is close to 0, two partitions are different from each other.

\subsection{Method for testing of algorithms}

We need to test our algorithm on different networks.
First one is the Zachary karate club network~\cite{%
	zachary1977Karate}.
It is a well known network dataset of a karate club where members of 
the club are divided into two groups after a dispute over lesson prices,
i.e. first group continued with the president of the club while the second group continued with the instructor. 
The network dataset has ground-truth community structure; 
so we can compare it with the communities identified by our algorithm.

As a second group of testing, we use real-life networks provided by 
SNAP~\cite{datasetSNAP2014}; namely DBLP dataset, 
Amazon co-purchase network, Youtube network and 
European-email network,
which have available ground-truth communities.
We run some of the known community detection algorithms,
which are 
Newman's fast greedy algorithm~\cite{%
	newman2004fast}, 
Infomap~\cite{%
	rosvall2007Infomap}, 
Louvain~\cite{%
	blondel2008Louvain} and 
Label Propagation (LPA)~\cite{%
	raghavan2007LPA}
on these network datasets 
and compare their results with the results of our algorithm.
We compare the partitions identified by each algorithm with partitions of ground-truth of these networks using NMI.
We also measure the execution times of all the algorithms
(we use a standard laptop computer having a 2.2 GHz Intel Core i7 processor with 4-cores).

When we do not have the ground-truth community structure, 
we can not benchmark the results of a newly proposed community detection algorithm.
One can try a comparative analysis by running a set of algorithms on a network dataset and 
make a pairwise comparison between identified partitions of these algorithms.
This is not a good way of quality testing for an algorithm 
because there is not a universally ``best'' community detection algorithm that can be used as 
gold standard.
We carried such a comparative analysis, however we could only see how close or far each algorithm to any other algorithm in terms of  number of identified communities or NMI values.
For that reason, without ground-truth, quality test of an algorithm in terms of finding correct communities can not be done.

Many real-life network datasets do not have ground-truth of community structure.
In such cases, 
computer generated networks like LFR benchmark networks~\cite{lancichinetti2008benchmark} 
with planted community structure can be helpful.
LFR algorithm generates networks with ground-truth community structure
using parameter vector of 
$[N, \langle k\rangle, k_{max}, C_{min}, C_{max}, \mu]$,
where 
$N$ is the number of nodes and
$\mu$ is the mixing parameter.
We investigate response of community detection algorithms to datasets generated with various mixing values.
As $\mu$ increases the community structure becomes more blurry and difficult to detect.
Being nondeterministic, LFR can generate different networks for the same parameter vector.
In order to avoid potential bias of an algorithm to a single network,
we generate 100 LFR networks for each vector and report the averages.

%
\section{Our Approach}

Our approach is based on building a preference network
where each connected component is declared as a community.
Given a network, we build its corresponding preference network
using the preference of each node for other nodes to be in same community with.
Every node prefers to be in the same community with certain nodes and
we simply try to satisfy these requests.
In this study we implement the case 
where each node is allowed to select only one node,
which is the most preferred node to be with.
It is relatively easy to extend this approach 
to nodes preferring two, three or more nodes, too.
First, we describe how to satisfy such requests 
by means of preference network.
Then we investigate ways to decide which node or nodes to be with.

\subsection{Preference network}

Let $G = (V, E)$ be an undirected network
where $V$ and $E$ are the sets of nodes and of edges, respectively.
Define a \emph{prefer} function $p \colon V \to V$ such that $p(i) = j$ 
iff
node $i$ prefers to be in the same community with node $j$.
If we connect $i$ to $p(i)$, 
clearly $p$ induces a new directed network on $V$,
but we will use the corresponding undirected network.
Using $p$, 
we define a new undirected network $G^{p} = (V, E^{p})$ such that 
nodes $i$ and $j$ are connected,
i.e. $(i, j) \in E^{p}$, 
iff
either $p(i) = j$ or $p(j) = i$.
We call this network as the \emph{preference network.}
We consider the components of $G^{p}$ as communities.
Hence we satisfy the rule that every node is in the same community with its preferred node.
Note that preference network is not a tree 
since it may have cycles
as in the case of 
node $a$ prefers node $b$,
$b$ prefers node $c$,
$c$ prefers $a$.
	See SI for an algorithm of 
	extracting communities using preference network~\cite{%
		Tasgin2017SI}.
%

\subsection{Deciding which node to be with}

Now we can investigate a selection method of preferred node.
First of all,
with the given definition,
there is nothing that restricts a node to prefer any other node in the network
even if the preferred node is not connected to the node.
For example a node could prefer the node with the largest betweenness centrality.
This view is too general and requires global information.

We restrict the selection of preferred node to the local neighborhood of every node.
We calculate a score for each node in the local neighborhood $A(i)$ around $i$, with respect to $i$.
Then select the node with highest score (detailed later) as preferred node.
That is, we define the function $p$ as;
\[
	p(i) = \argmax {j \in A(i)} s_{i}(j)
\]
where
$s_{i}(j)$ being the score of node $j$ with respect to $i$.
In tie situations, 
i.e. when two or more neighbors having the highest score, 
node selects one of them randomly.
Note that
the score $s_{i}(j)$ of $j$ depends on the node $i$.
The score can be interpreted as a measure of how ``important'' is node $j$ for $i$.
If node $j$ is the only connection of $i$,
it has to have very big value.
If $i$ has many neighbors,
then $j$ may not be very important for $i$.
Hence the very same node $j$ usually has different scores with respect to some other nodes, 
i.e. $s_{i}(j) \ne s_{k}(j)$.

Local neighborhood can be defined in a number of ways.
One may define it as
the nodes whose distance is not more than $\ell$ to $i$.
It may be the nodes whose distances to $i$ are exactly $\ell$.
We may also include node $i$ itself to the local neighborhood.
In this case $i$ may prefer to be in the same community with itself.
For this study, 
we take $A(i)$ as the 1-neighborhood of $i$,
i.e.
the set of nodes whose distance to $i$ is exactly 1, which is denoted by $\Gamma(i)$.
%

\subsection{Candidates of score metric}

There are a number of candidates for score $s_{i}(j)$ calculation of $j$ with respect to $i$.\\
(i)~The simplest one is to assign a random number for each neighbor of $i$ as its score.
Probably this is not a good choice,
since random function will decide independent of node $j$ and node $i$; and their relations with each other.
(ii)~Nodes with more connections are usually considered to be more important in a network.
So, 
as a second choice, 
we can use the degree of the nodes, 
i.e. $s_{i}(j) = \hbAbs{\Gamma(j)}$.\\
The degree 
of node $j$ is also independent of $i$.
So we do not incorporate what $i$ thinks of $j$ in the score $s_{i}(j)$.\\
(iii)~A third candidate is clustering coefficient of $j$,
which is an indication of how densely connected its immediate neighborhood, 
i.e. $s_{i}(j) = CC_j$.
This is again a value, 
which does not directly depend on $i$ but may have a meaning to $j$ and $i$, 
i.e. there is a chance that high clustering coefficient of $j$ is 
at least partly because of triangles shared by $j$ and $i$. \\
(iv)~Having common neighbors is an important feature in social networks.
From the definition of community, members inside the community should have more edges among themselves which leads to more common neighbors of nodes inside a community.
Number of common neighbors $i$ and $j$ is the number of triangles having $i$ and $j$ as two corners.
For this reason, as a fourth candidate we can use the number of common neighbors of $i$ and $j$,
i.e., 
$s_{i}(j)=\hbAbs{\Gamma(i) \cap \Gamma(j)}$.\\
(v)~As a fifth candidate,
we can use spread capability of a neighbor $j$ around node $i$,
i.e. $s_{i}(j)=\sigma_{i,j}$.
It both contains number of common neighbors and triangle cascades as discussed earlier.\\
(vi)~And as the sixth candidate, we can use Jaccard similarity of $i$ and $j$ as score, 
i.e. $s_{i}(j) = J(i,j)$.\\

\section{Results and Discussion}

\subsection{Selection of best score metric}

We first analyze alternative score metrics in our algorithm and try to find which one 
performs better in community detection.
We run our algorithm on generated LFR networks~\cite{lancichinetti2008benchmark} of 
$1,000$ nodes using all the score metrics, $s_{i}(j)$, as the method of preferred node selection.
NMI values and execution times are measured. 
The results of our algorithm using six different score metrics on LFR networks 
generated with increasing mixing values ($\mu$) are in 
\reffig{fig:LFR1000_Networks}.
We observe that number of common neighbors 
is the best score metric among six alternatives; it has the best NMI values and 
can identify exact community structure on networks generated with
 $\mu=0.1$ and $\mu=0.2$.
Spread capability score metric has the second best results; it is better than Jaccard similarity
where Jaccard similarity metric finds 4-5 times more number of 
communities compared to ground-truth of LFR networks. 
Other three metrics; namely 
random score assignment, 
degree and 
clustering coefficient 
can identify communities to a degree but not as successful as the ones mentioned above.
In the second group of metrics, clustering coefficient is the best one and 
our algorithm using clustering coefficient score can find communities on networks generated with low $\mu$.
Interestingly, random score metric can identify a group of 
communities successfully on these networks.
In general, our algorithm using random score and degree based score metrics find 
less number of communities compared to ground-truth.

\newcommand{\myPlotFactor}{.23}
\begin{figure}
	\centering
	\subfloat[\label{fig:LFR1000_weightMethods}]{%
		\includegraphics
			[scale=\myPlotFactor]
			{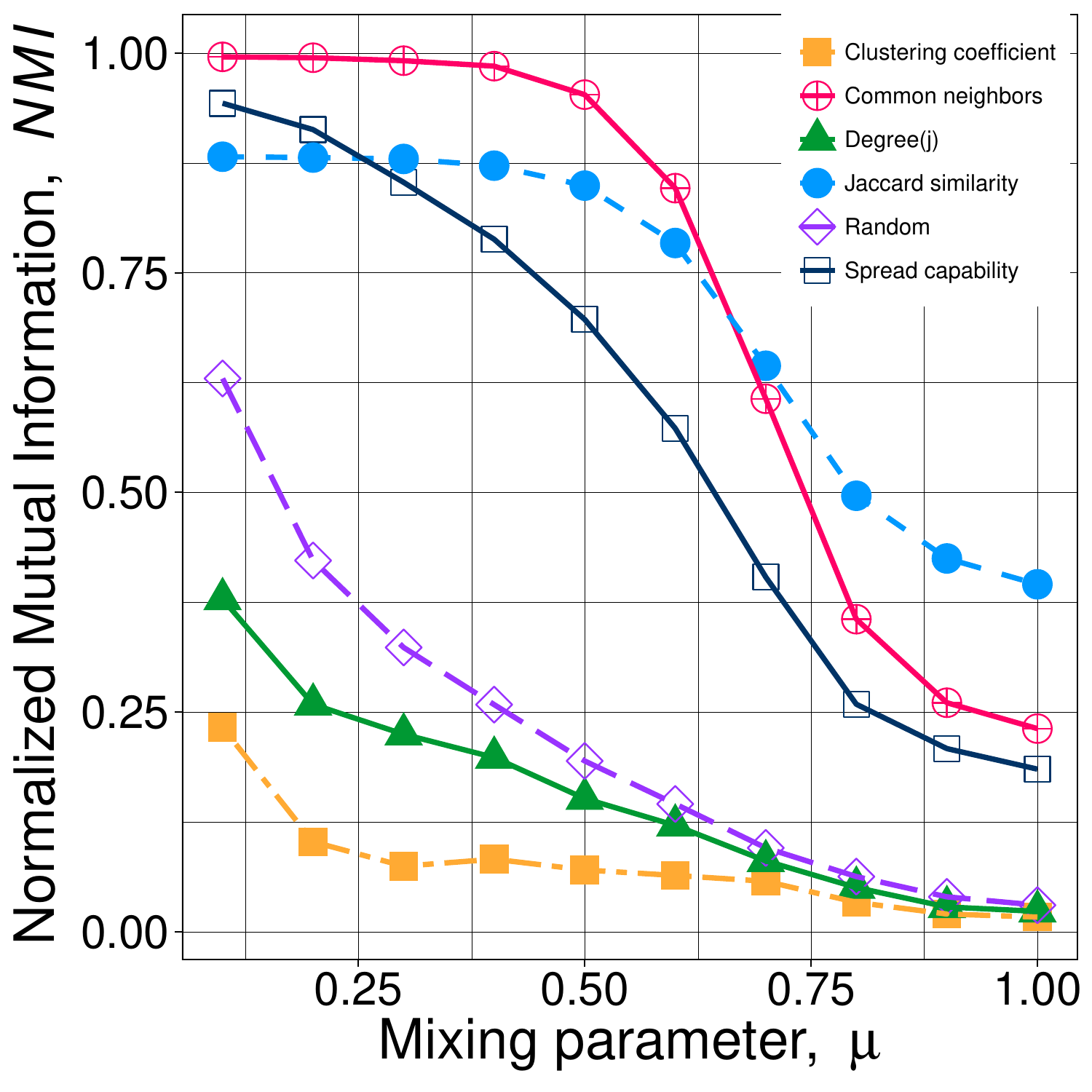}
	}
	\subfloat[\label{fig:LFR1000_executionTimes}]{
		\includegraphics
			[scale=\myPlotFactor]
			{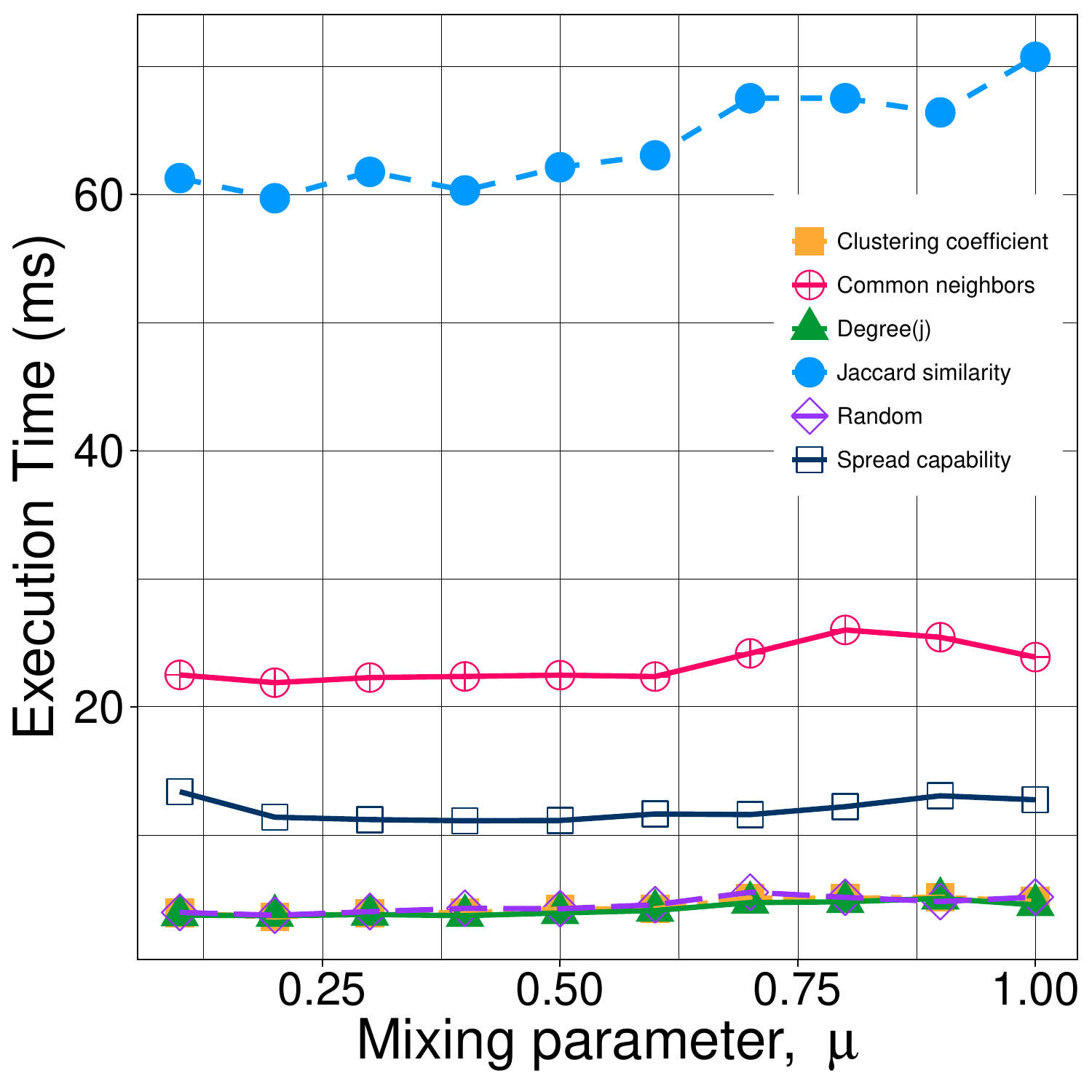}
	}

	\caption{
		NMI comparison and execution times of different score metrics on 
		LFR benchmark network datasets. 
		$N$=$1,000$, $\langle k\rangle$=15, $k_{max}$=50, $C_{min}$=10, $C_{max}$=50
	}
	\label{fig:LFR1000_Networks}
\end{figure} 

It is trivial that calculation of simple score metrics requires less computation time compared to calculation of other score metrics,
i.e. execution time of our algorithm using random score, 
degree and clustering coefficient all have less execution times as given in \reffig{fig:LFR1000_executionTimes}.
On the other hand, calculation of common neighbors, spread capability and Jaccard similarity require more 
computation time, as these metrics are calculated for each pair of nodes (i.e. number of edges),
however these metrics have better results in terms of community detection.
Hence, we select the two best performing score metrics for our algorithm, 
namely, common neighbors and spread capability,
denoted as \hbAlgorithmCN\ and \hbAlgorithmSC, 
respectively.
We use these score metrics in our algorithm for 
comparative analysis with other known algorithms on generated and real-life networks.

\subsection{Results on networks}
\subsubsection{Zachary karate club network}

\begin{figure}
	\begin{center}
		\includegraphics[width=1\linewidth]
			{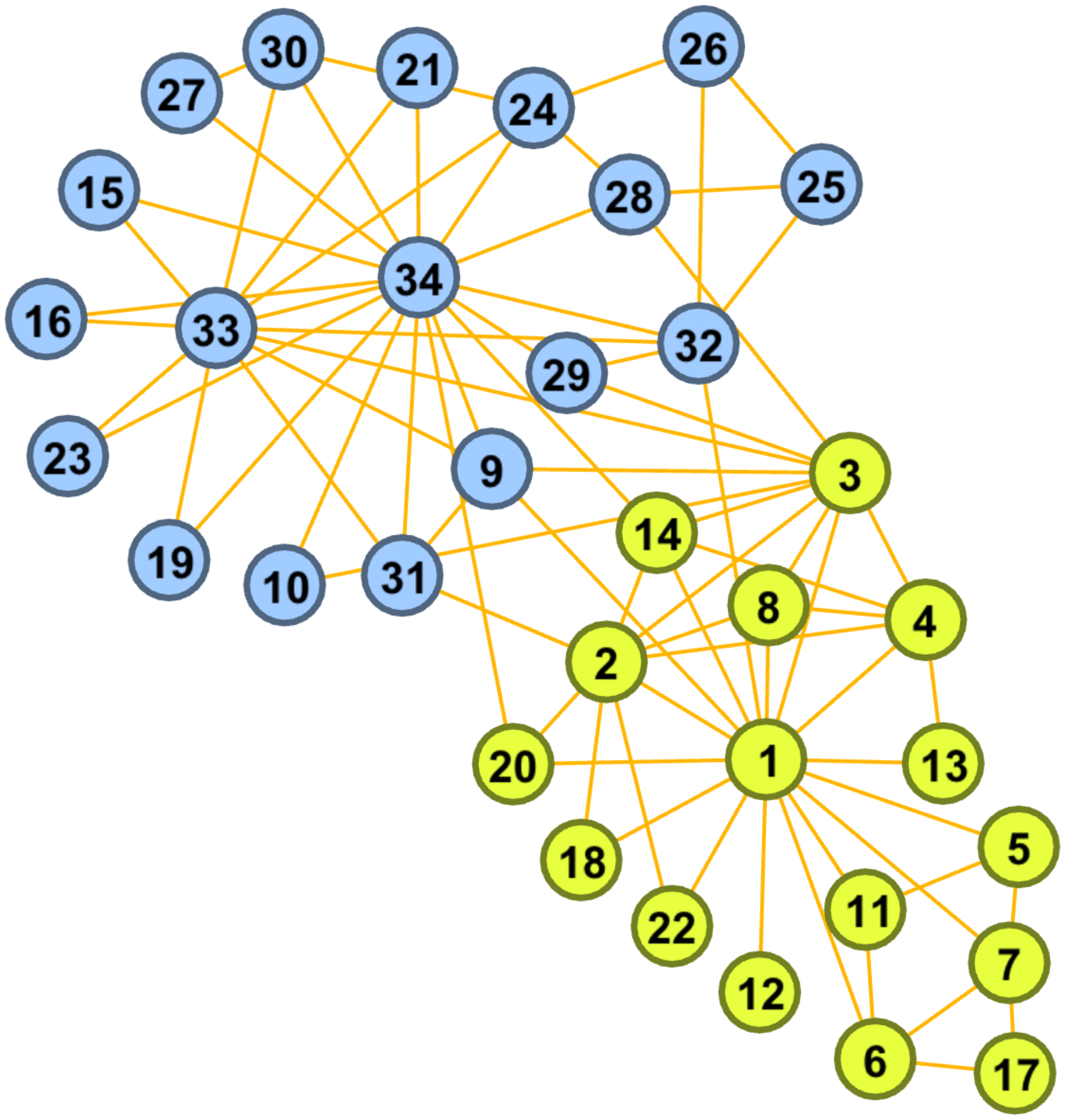}
		\caption{Zachary karate club: Identified communities by our \hbAlgorithmCN\ algorithm}
		\label{fig:karateClubCommunitiesPCN}
	\end{center}
\end{figure}

We run our algorithm on Zachary karate club network and 
compare the identified communities with those of ground-truth.
Our algorithm with common neighbors score metric, namely \hbAlgorithmCN, 
identifies two communities as seen in \reffig{fig:karateClubCommunitiesPCN}.
Only node 9 is misidentified by our algorithm.
Node 9 actually has more connections in its identified community and
the ground-truth metadata may not reflect the actual community.
All the other nodes are identified correctly.

\subsubsection{Large real-life networks}

We run our algorithm with both score metrics, \hbAlgorithmCN\  and \hbAlgorithmSC, on
large networks with ground-truth 
communities provided by SNAP~\cite{datasetSNAP2014}. 
For comparative analysis, Infomap, Louvain, LPA and 
Newman's fast greedy algorithm are also run on these networks.
We omit Newman's algorithm on Youtube network dataset
since it could not finish due to long execution time.
Results are presented in \reftbl{tbl:tableLargeNetworks}.
On all of the four real-life networks, number of communities found 
by \hbAlgorithmCN, \hbAlgorithmSC, Infomap and LPA are close to each other 
and not far from the ground-truth (one exception is the Youtube network).
On all of these networks, our algorithm finds more number of communities because of its local nature.

In general, performance of Louvain and Newman's algorithm on large real-life networks is low.
Their NMI scores are very low and number of identified communities by these algorithms are far from those of ground-truth.
They find very few number of communities compared to ground-truth.
These two algorithms perform better on European-email network.
Infomap and LPA algorithms generally have better NMI values compared to other algorithms,
however LPA finds only two communities in European-email network where there are 42 ground-truth communities.
Our algorithm, with both score metrics (\hbAlgorithmCN\ and \hbAlgorithmSC),
performs well on most of the networks with good NMI values. 
However it performs poorly on Youtube network where all the other algorithms have similar bad results.
This may be due to very small clustering coefficient of Youtube network, 
i.e. no trivial community structure is available.

\begin{table*}[htbp]
	\caption{
		Large real-life networks 
	}
	\begin{center} 
	\scalebox{0.63}{
		\begin{tabular}{|l|r|r|r|r|r|r|r|r|r|r|r|r|r|r|r|r|r|r|r|r|r|r|}
			\hline
			\toprule	
			\multirow{2}{*}{Network} &	\multirow{2}{*}{$|V|$} &\multirow{2}{*}{$|E|$}& \multirow{2}{*}{CC} &  \multicolumn{7}{c}{\# communities} & \multicolumn{6}{|c|}{NMI}& \multicolumn{6}{c|}{execution time (ms)}\\ \cline{5-23}
			{} & {} & {} & {} 				& GT		& \hbAlgorithmCN   &  \hbAlgorithmSC	& Inf 		& LPA 	&   Lvn  	& NM 	& \hbAlgorithmCN	&	\hbAlgorithmSC	&  Inf 	&	LPA	&Lvn		&NM	& \hbAlgorithmCN	&	\hbAlgorithmSC	&  Inf 	&	LPA	&Lvn		&NM	\\ \hline	                               
European-email	&1,005& 	16,064	&	0.40	&	42	&	35		&32		&	38	&2		&	25	&28		&0.34& 0.17&0.62&0.01&0.54&0.46 & 192 & 133 & 133 & 40 & 69 & 187\\
DBLP	&   	317,080 	&	1,049,866	&	0.63 	&13,477	&	28,799	& 28,798	&30,811	&36,291 	&	565	&3,165	&0.58& 0.57&0.65&0.64&0.13&	0.16 & 4,652 & 3,879 & 35,753 & 106,410 & 8,217 & 4,362,272\\
Amazon	&	334,863	&	925,872	&	0.40	& 75,149	&	36,514	& 36,519 	&35,139	&23,869 	&	248	&1,474	&0.58& 0.59&0.60&0.54&0.11&	0.11 & 2,911 & 3,453 & 43,253 & 83,532 & 8,017 & 1,422,590\\
Youtube	&	1,134,890	&	2,987,624	&	0.08	& 8,385	&	78,021	& 78,053 	&102,125	&83,256  	&9,616	&	-	&0.07& 0.08&0.13&0.07&0.06&	-       & 105,528 & 421,593 & 188,037 & 1,362,241 & 52,798 & -\\
			\bottomrule
			\hline
		\end{tabular}
		}
	\end{center}
	\label{tbl:tableLargeNetworks}

	\scalebox{0.65}{
	\begin{minipage}{0.85\textwidth}%
		\begin{flushleft}
	  		\small GT : Ground-truth \\%
			\small \hbAlgorithmCN : Community detection using preference network with common neighbors score\\%
			\small \hbAlgorithmSC : Community detection using preference network with spread capability score\\%
			\small Inf     : Infomap algorithm\\%
			\small LPA  : Label propagation algorithm\\%
			\small Lvn   : Louvain community detection algorithm\\%
			\small NM   : Newman's fast algorithm\\%
		\end{flushleft}
	\end{minipage}%
	}
\end{table*}

%
\subsubsection{Generated networks}
We perform the similar comparative analysis on generated LFR networks of $1,000$ and $5,000$ nodes
as reported in \reffig{fig:LFR_knownMethods} and \reffig{fig:LFR_knownMethodsLarge}, 
respectively.
We present the detailed results of algorithms on LFR networks of $5,000$ nodes in \reftbl{tbl:tableLFRLarge}.
As described earlier, we generate 100 LFR networks per $\mu$ value,
run the algorithms on all 100 generated datasets and
averaged the results for each algorithm.
On LFR networks with $1,000$ nodes, our algorithm with common neighbors (\hbAlgorithmCN) 
is among the top 3 best performing algorithms 
according to the NMI values; 
on most of the networks, Infomap and our algorithm find the 
best results and LPA is in the third place.
Our algorithm with spread capability (\hbAlgorithmSC) has lower NMI values but still performs better than Newman's algorithm.
On the networks generated with higher mixing values (i.e. $\mu > 0.5$),
Infomap and LPA tend to find small number of communities and 
sometimes they group all the nodes into a single community.
Louvain and Newman's fast algorithm also find very few number of communities on these networks.
However our algorithm, with both of the score metrics, can still find communities successfully. 
NMI values of our algorithm are better than those of other algorithms and number of communities found by 
our method do not differ much from the ground-truth compared to other algorithms.

\newcommand{\myPlotFactorB}{.40}
\begin{figure*}
	\centering
	\subfloat[\label{fig:LFR_knownMethods}]{%
		\includegraphics
			[scale=\myPlotFactorB]
			{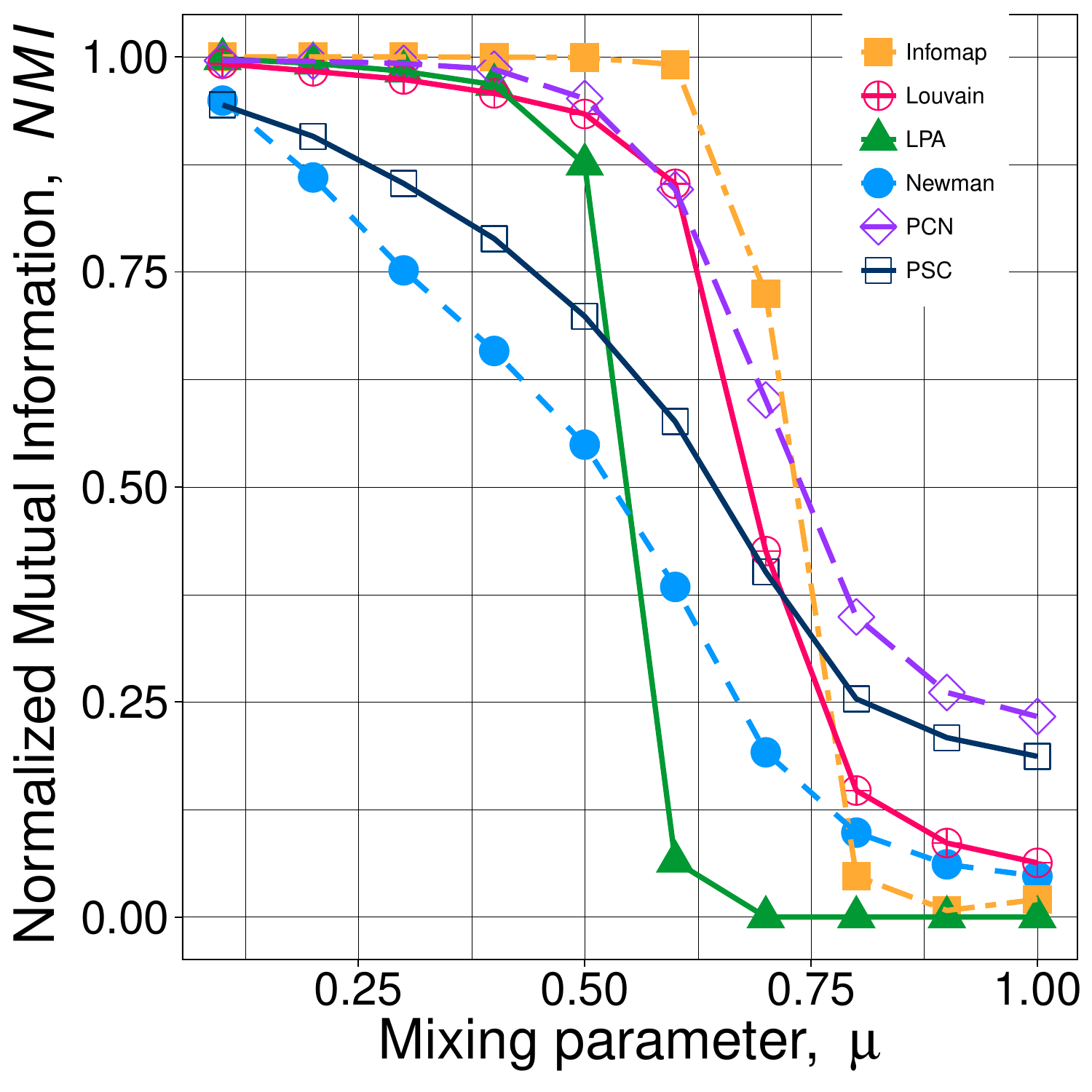}
	}
	\subfloat[\label{fig:LFRMedium_executionTimes}]{
		\includegraphics
			[scale=\myPlotFactorB]
			{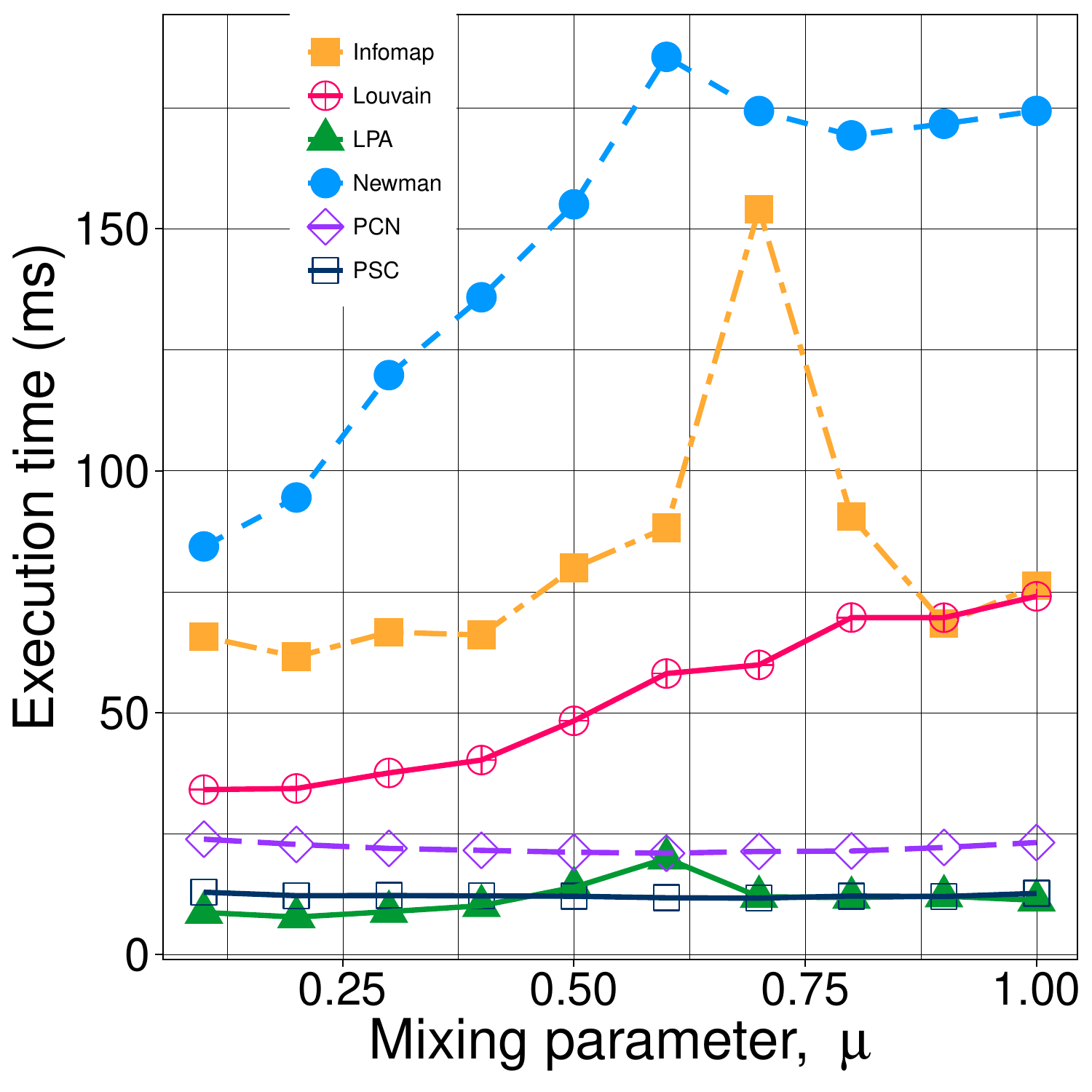}
	}
	\caption{
		Comparison of our method and known algorithms on 
		LFR benchmark network datasets (NMI and execution times). 
		$N$=$1,000$, $\langle k\rangle$=15, $k_{max}$=50, $C_{min}$=10, $C_{max}$=50
	}
	\label{fig:LFR1000_ComparativeAnalysis}
\end{figure*} 

\newcommand{\myPlotFactorC}{.40}
\begin{figure*}
	\centering
	\subfloat[\label{fig:LFR_knownMethodsLarge}]{%
		\includegraphics
			[scale=\myPlotFactorC]
			{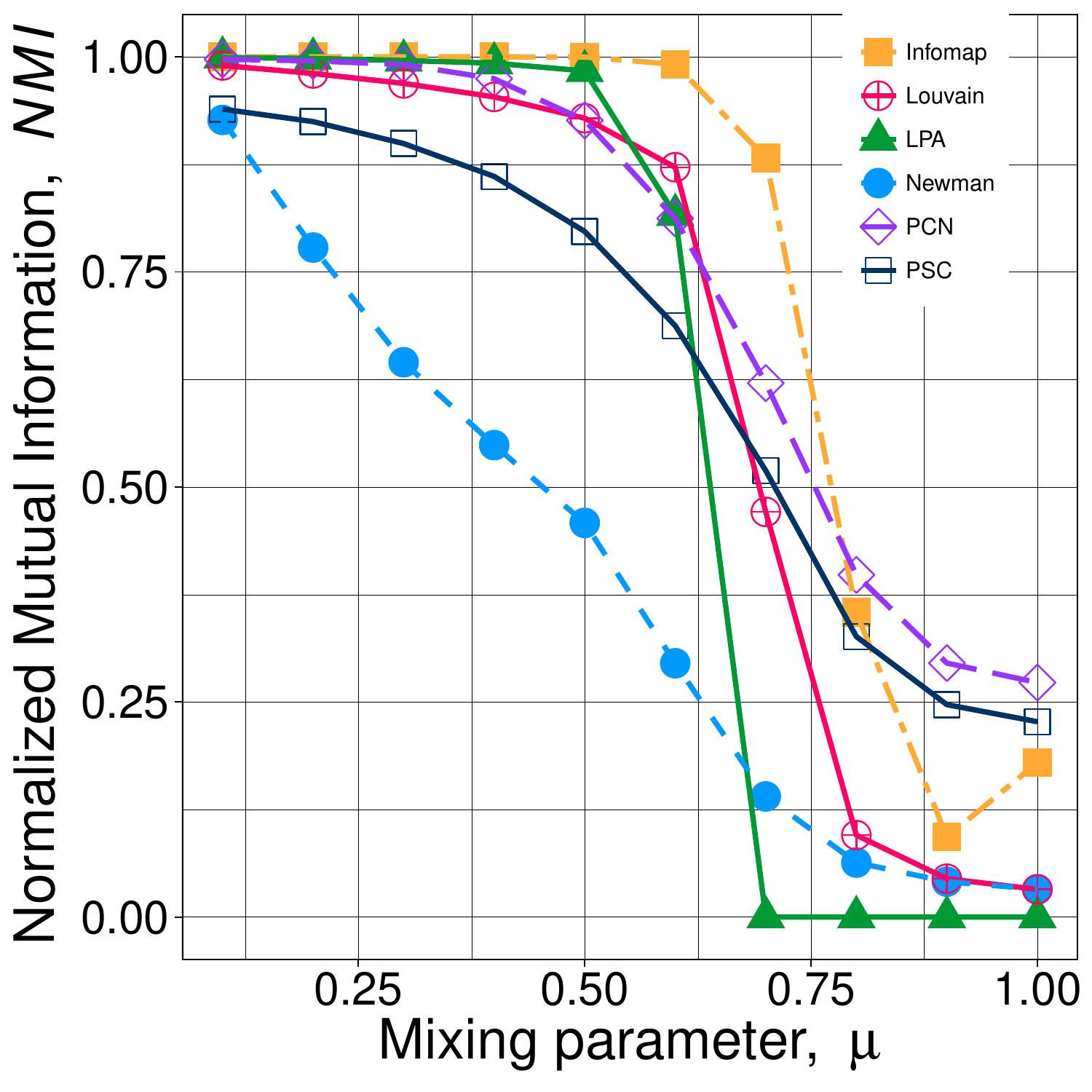}
	}
	\subfloat[\label{fig:LFRLarge_executionTimes}]{
		\includegraphics
			[scale=\myPlotFactorC]
			{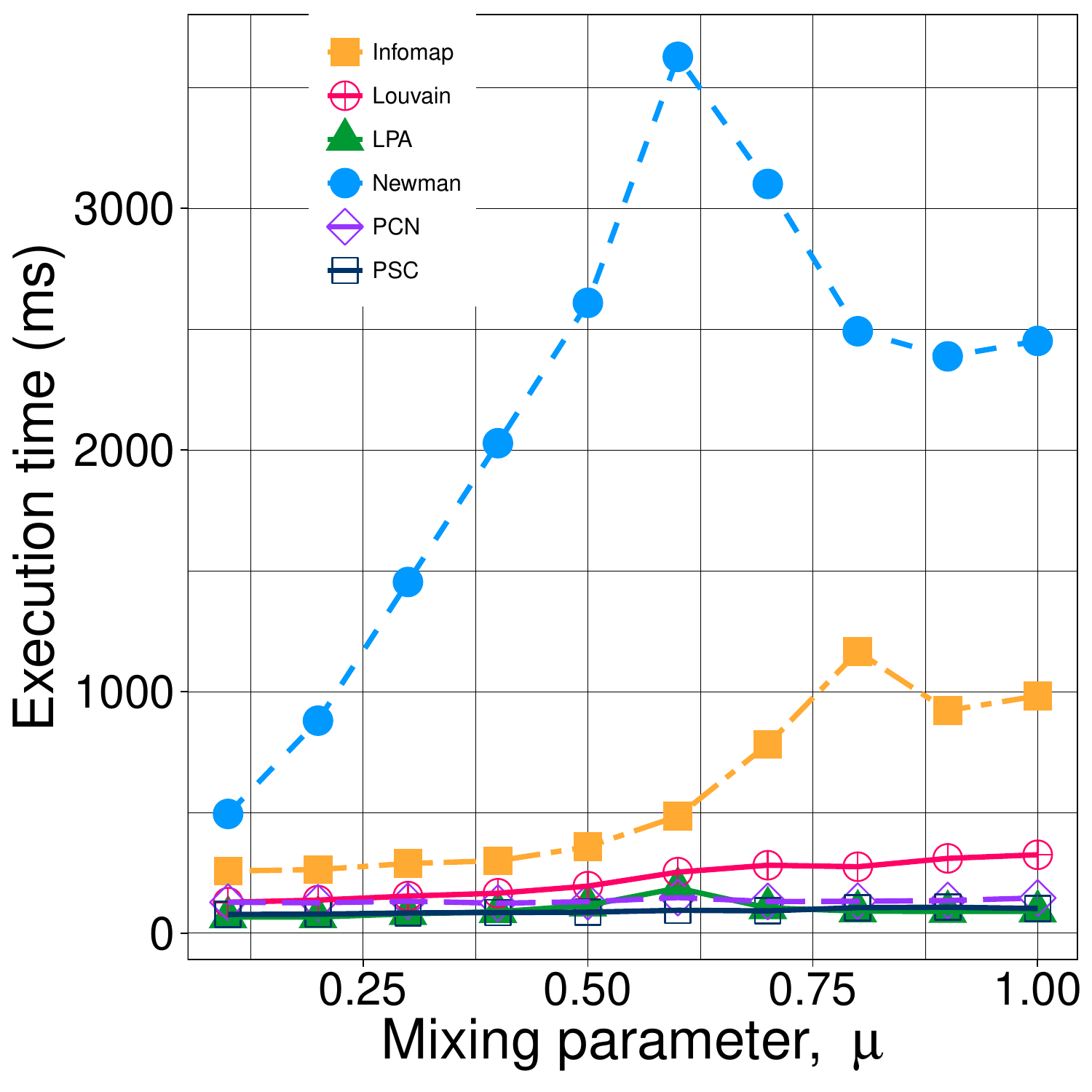}
	}
	\caption{
		Comparison of our method and known algorithms on 
		LFR benchmark network datasets (NMI and execution times). 
		$N$=$5,000$, $\langle k\rangle$=15, $k_{max}$=75, $C_{min}$=20, $C_{max}$=100
	}
	\label{fig:LFR5000_ComparativeAnalysis}
\end{figure*} 

On LFR networks of $5,000$ nodes, our algorithm has better results
compared to its performance on previous set of LFR networks of $1,000$ nodes.
However with spread capability score metric, it finds more granular communities, 
which leads to greater number of communities compared to ground-truth.
Newman's algorithm and Louvain algorithm find very few number of communities; 
they tend to merge communities which may lead to resolution limit~\cite{fortunato2007resolution}.

Infomap and LPA are both successful on large networks when mixing parameter is low, 
however their quality degrades with increasing mixing parameter where our algorithm can still identify communities successfully.
 
One of the main differences between our algorithm and LPA is that, we do not assign community labels
to nodes but keep the information of who prefers whom to be in same community using a preference network.
During execution steps of LPA, a node updates its community label according to majority of labels of its neighbors.
However when all or part of those neighbors update their labels to be in a different community, 
then the node will fall apart from them, it will be in a different community 
(however it wanted to be in same community with them and updated its label accordingly).
Using a preference network, we preserve all the preferences made by 
each node throughout the execution of algorithm (because we do not update any label).
And aggregation of all these preferences will eventually lead to a good community structure.

\subsection{Performance of the algorithm}

In this section we discuss the performance and time-complexity of our algorithm.
The details are given in SI~\cite{%
	Tasgin2017SI}.
We use common neighbors as edge weights for \hbAlgorithmCN\ 
and spread capability for \hbAlgorithmSC.
Given a network $G = (V, E)$, 
where 
maximum degree of nodes is $k_{\mathrm{max}}$,
calculating edge weights 
has $O(\hbAbs{E})$
and $O(\hbAbs{V})$ time-complexity, 
for  \hbAlgorithmCN\   and \hbAlgorithmSC\ respectively.
As obtaining communities on preference network requires $O(\hbAbs{V})$ time-complexity,
the overall time-complexity of our algorithm is $O(\hbAbs{E} + \hbAbs{V})$ for \hbAlgorithmCN\
and $O(\hbAbs{V} + \hbAbs{V})=O(\hbAbs{V})$ for \hbAlgorithmSC.

Our algorithm is fast and suitable for very large networks on a single processor environment
where calculations are done in a sequential manner.
Its speed can be improved further by parallel execution.
As multiprocessors become available,
how well an algorithm can be distributed over parallel processors becomes an important topic.
Our algorithm requires calculations such as number of common neighbors or spread capability,
which can be related to edges around a node.
It can be distributed to as many as $N$ processors easily.
In this case each processor handles the local calculations around each node in the network.
Then each node will have scores to 
decide its preferred node in such a parallel and fast way.
Network data can be redundantly replicated to all processors
in order to avoid the computation overhead of data splitting process.
Note that obtaining components is not easily distributed in a parallel fashion since 
component discovery in preference network can not be split into independent tasks.
However this step needs fewer steps of computation and has less impact on overall performance.

\begin{table*}[htbp]
\centering
\caption{Generated LFR benchmark networks of 5000 nodes  }
\label{tbl:tableLFRLarge}
\scalebox{0.73}{
\begin{tabular}{|l|r|r||r|r|r|r|r|r|r|r|r|r|r|r|r|r|r|r|r|r|r|r|r|}
\hline
\multicolumn{1}{|c|}{\multirow{2}{*}{Network}} & \multicolumn{1}{c|}{\multirow{2}{*}{$|V|$}} & \multicolumn{1}{c|}{\multirow{2}{*}{$\mu$}} & \multicolumn{1}{c|}{\multirow{2}{*}{$|E|$}} & \multicolumn{1}{c|}{\multirow{2}{*}{CC}} & \multicolumn{7}{c|}{\# communities}                                                                                                                                                       & \multicolumn{6}{c|}{NMI}                                                                                                                                        & \multicolumn{6}{c|}{execution time (ms)}                                                                                                                        \\ \cline{6-24} 
\multicolumn{1}{|c|}{}                         & \multicolumn{1}{c|}{}                     & \multicolumn{1}{c|}{}                        & \multicolumn{1}{c|}{}                     & \multicolumn{1}{c|}{}                    & \multicolumn{1}{c|}{GT} & \multicolumn{1}{c|}{PCN} & \multicolumn{1}{c|}{PSC} & \multicolumn{1}{c|}{Inf} & \multicolumn{1}{c|}{LPA} & \multicolumn{1}{c|}{Lvn} & \multicolumn{1}{c|}{NM} & \multicolumn{1}{c|}{PCN} & \multicolumn{1}{c|}{PSC} & \multicolumn{1}{c|}{Inf} & \multicolumn{1}{c|}{LPA} & \multicolumn{1}{c|}{NM} & \multicolumn{1}{c|}{Lvn} & \multicolumn{1}{c|}{PCN} & \multicolumn{1}{c|}{PSC} & \multicolumn{1}{c|}{Inf} & \multicolumn{1}{c|}{LPA} & \multicolumn{1}{c|}{Lvn} & \multicolumn{1}{c|}{NM} \\ \hline
LFR-1                                          & 5,000                                     & 0.1                                          & 38,868                                    & 0.51                                     & 101                     & 105                      & 240                      & 101                      & 100                      & 89                       & 64                       & 0.99                     & 0.94                     & 0.99                     & 0.99                     & 0.93                     & 0.99                     & 127                      & 77                       & 256                      & 66                       & 127                      & 493                      \\ \hline
LFR-2                                          & 5,000                                     & 0.2                                          & 38,955                                    & 0.37                                     & 101                     & 108                      & 254                      & 101                      & 100                      & 81                       & 31                       & 0.99                     & 0.92                     & 0.99                     & 0.99                     & 0.78                     & 0.98                     & 125                      & 78                       & 263                      & 66                       & 136                      & 879                      \\ \hline
LFR-3                                          & 5,000                                     & 0.3                                          & 38,871                                    & 0.25                                     & 101                     & 111                      & 266                      & 101                      & 98                       & 73                       & 18                       & 0.99                     & 0.90                     & 0.99                     & 0.99                     & 0.64                     & 0.97                     & 132                      & 83                       & 288                      & 79                       & 153                      & 1,453                    \\ \hline
LFR-4                                          & 5,000                                     & 0.4                                          & 38,930                                    & 0.16                                     & 101                     & 121                      & 283                      & 101                      & 96                       & 64                       & 12                       & 0.97                     & 0.86                     & 0.99                     & 0.99                     & 0.55                     & 0.95                     & 123                      & 85                       & 299                      & 88                       & 165                      & 2,028                    \\ \hline
LFR-5                                          & 5,000                                     & 0.5                                          & 38,973                                    & 0.10                                     & 100                     & 142                      & 294                      & 100                      & 91                       & 53                       & 9                        & 0.93                     & 0.80                     & 0.99                     & 0.98                     & 0.46                     & 0.93                     & 130                      & 86                       & 357                      & 115                      & 194                      & 2,609                    \\ \hline
LFR-6                                          & 5,000                                     & 0.6                                          & 38,973                                    & 0.05                                     & 100                     & 185                      & 300                      & 103                      & 74                       & 41                       & 11                       & 0.81                     & 0.69                     & 0.99                     & 0.81                     & 0.30                     & 0.87                     & 147                      & 94                       & 483                      & 186                      & 252                      & 3,628                    \\ \hline
LFR-7                                          & 5,000                                     & 0.7                                          & 38,969                                    & 0.02                                     & 101                     & 243                      & 280                      & 159                      & 1                        & 25                       & 14                       & 0.62                     & 0.52                     & 0.88                     & 0.00                     & 0.14                     & 0.47                     & 130                      & 92                       & 781                      & 104                      & 281                      & 3,101                    \\ \hline
LFR-8                                          & 5,000                                     & 0.8                                          & 38,923                                    & 0.01                                     & 100                     & 269                      & 243                      & 227                      & 1                        & 12                       & 13                       & 0.40                     & 0.33                     & 0.35                     & 0.00                     & 0.06                     & 0.10                     & 131                      & 105                      & 1,165                     & 91                       & 274                      & 2,491                    \\ \hline
LFR-9                                          & 5,000                                     & 0.9                                          & 38,986                                    & 0.01                                     & 102                     & 278                      & 238                      & 76                       & 1                        & 12                       & 13                       & 0.30                     & 0.25                     & 0.09                     & 0.00                     & 0.04                     & 0.04                     & 134                      & 107                      & 921                      & 89                       & 309                      & 2,388                    \\ \hline
LFR-10                                         & 5,000                                     & 1.0                                          & 38,947                                    & 0.01                                     & 101                     & 285                      & 242                      & 81                       & 1                        & 12                       & 13                       & 0.27                     & 0.23                     & 0.18                     & 0.00                     & 0.03                     & 0.03                     & 145                      & 102                      & 982                      & 90                       & 323                      & 2,451                    \\ \hline
\end{tabular}
}

\end{table*}

\section{Conclusion}

We propose a new local community detection algorithm with two variants, 
 \hbAlgorithmCN\ and \hbAlgorithmSC, which builds a preference network using two 
different node similarity score metrics; namely 
common neighbors and gossip spread capability.
Although it uses only local information, its performance is good
especially when community structure is not easily detectable. 
On LFR networks generated with higher mixing value ($\mu>0.7$)
our algorithm performs better than all the other algorithms used for comparison in this paper.
On such networks,
algorithms like 
Infomap and LPA merge all the nodes into a single community, 
where these algorithms are stuck in a local optimum and fail to identify communities.
Our algorithm identifies communities in many large real-life networks
with high accuracy in a fast way.

We think that building a preference network to identify communities is a simple and powerful approach.
With this, we preserve the preferences of all of the nodes for being in same community with another node, whether they are highly connected or have a few connections.
This approach prevents loss of granular community information especially in very large networks.
Even with random score assignment, 
our algorithm can identify many communities on generated LFR networks.

Due to its local nature, our algorithm is scalable and fast,
i.e. it needs only a single pass on the whole network to construct a preference network
and similarity metric used for this construction can 
be evaluated in 1-neighborhood of each node.
It can run on very large networks without loss of quality and performance.

We haven't implemented a distributed or parallel version of our algorithm 
however  it is suitable for parallel processing in a distributed environment.
It can be deployed as agents on different parts of a large real-life network
which is evolving over time.
On such a network, collecting the data of the whole network is costly (time, space, computation), 
while information about small parts of the network can easily be obtained and analyzed by 
each nearby agent for community detection.
Agents can easily identify community structures of that particular area without 
knowing the rest of the network;
which is a valuable information at that scale
and can be used in real-time by systems like peer-to-peer networks.\\

Source code of our algorithm is available online at:\\
\href{https://github.com/murselTasginBoun/CDPN}
{https://github.com/murselTasginBoun/CDPN}

\section*{Acknowledgments}
Thanks to Mark Newman, Vincent Blondel and Martin Rosvall for the source codes of their community detection algorithms. 
Thanks to Mark Newman, Jure Leskovec and Vladimir Batagelj for the network datasets used.

This work was partially supported 
by the Turkish State Planning Organization (DPT) TAM Project (2007K120610).


\section{Supplementary Information}

The performance of the algorithm is discussed in two ways.
The single processor approach deals with $O(\cdot)$ complexity.
Another possibility is the scalability of the algorithm to multiprocessor running in parallel.

\subsection{Complexity on single processor}

In this section we approximate the time-complexity of the algorithm for real-life networks,
which are sparse networks.
The algorithm is composed of three steps.
(i)~For every edge $(i, j)$ in the network,
we assign a weight $w_{i j}$.
We use two metrics for $w_{i j}$,
namely number of common neighbors and spread capability.
(ii)~Then we construct a directed network,
where each node is connected to exactly one neighbor,
for which the weight is maximum.
(iii)~Finally,
we group the nodes into communities using the directed network.
First we investigate the complexity of calculating edge weights.
Then the complexity of obtaining communities from the directed network is investigated.

\subsubsection{Complexity of obtaining common neighbors}
\label{sec:ComplexityCommonNeighborsV4}

Let $(i,j) \in E$ be the edge connecting $i$ and $j$, 
and
 $k_{i}$, $k_{j}$ be the degrees of $i$ and $j$,
 respectively.
In order to find out 
if a neighbor $\ell$ of $i$ is also a neighbor of $j$,
we need to search $\ell$ in the neighbors of $j$.
Comparing $\ell$ with each neighbor of $j$ would require $k_{j}$ comparisons.
If we keep the neighbors of $j$ in a hash,
which provides direct access,
then the complexity of searching of $\ell$ in the hash would be $O(1)$.
Since there are $k_{i}$ neighbors of $i$, 
finding the common neighbors of $i$ and $j$ requires $k_{i}$ searches in the hash.
Note that if $k_{i} \ge k_{j}$,
then it is better to search neighbors of $j$ in the neighbors of $i$ in this situation.
Then the complexity finding the common neighbors of $i$ and $j$ is 
$O( \min\{ k_{i}, k_{j} \})$.
This is the complexity of calculating the weight of a single edge $(i, j)$ in network. 
Then the total number of comparisons required for all common neighbors can be obtained 
if we consider all the edges.
That is,
\begin{align*}
	\sum_{(i, j) \in E}  
		\min\{ k_{i}, k_{j} \}
	&<  \sum_{(i, j) \in E}  
			\min\{ \kMax, \kMax \}\\
	&<  \sum_{(i, j) \in E}  
			\kMax \\
	&<  \kMax 
		\sum_{(i, j) \in E}  
			1 \\
	&<  \kMax \hbAbs{E}
\end{align*}
where
$\kMax$ is the maximum degree in the network.
We get the worst case complexity of $O(\hbAbs{V}^{3})$ if the network is a complete graph,
where we have 
$\kMax = \hbAbs{V} - 1$ and
$\hbAbs{E} = {\hbAbs{V} \choose 2} \approx \hbAbs{V}^{2}$.
This is not a problem for a community detection algorithm,
since there is no community structure in a complete graph.
Fortunately, real life large networks are far from complete graphs.
Although the number of nodes is very large,
real-life networks are highly sparse,
i.e.,
$\hbAbs{E} \ll \hbAbs{V}^{2}$, 
and
their nodes are connected to a very small fraction of the nodes,
i.e.,
$\kMax \ll \hbAbs{V}$.
Note also that for networks with power-law degree distribution,
$\kMax$ is extremely high compared to degree of majority of the nodes.
So for real networks we can consider $\kMax$ as constant,
and the complexity becomes $O(\hbAbs{E})$.

\subsubsection{Complexity of obtaining spread capability}

Suppose we want to calculate the spread capability $\sigma_{i, j}$ of gossip originator $j$ around victim $i$,
which is given as
$
	\sigma_{i, j}
	= 
		{\hbAbs{\Gamma_{j}(i)}}
		/
		{\hbAbs{\Gamma(i)}}
$.
The denominator $\hbAbs{\Gamma(i)}$ is simply the degree $k_{i}$ of the victim node $i$.
The numerator $\hbAbs{\Gamma_{j}(i)}$ needs to be calculated.
We use common neighbor algorithm to obtain spread capability as follows.
In the first wave,
all common neighbors of $i$ and $j$ will receive the gossip from $j$.
If node $\ell$ is in the common neighbors of $i$ and $j$,
it will receive the gossip.
Now $\ell$ starts its wave,
i.e. a triangular cascade,
which passes gossip to all the nodes in the common neighbors of $i$ and $\ell$.
Any node that receives gossip,
will start its own wave.
As seen, 
we repeatedly use of common neighbors algorithm to propagate gossip from one node to another.

\newcommand{\myFigFactorA}{1}
\begin{figure}
	\centering
	\includegraphics[scale=\myFigFactorA]%
		{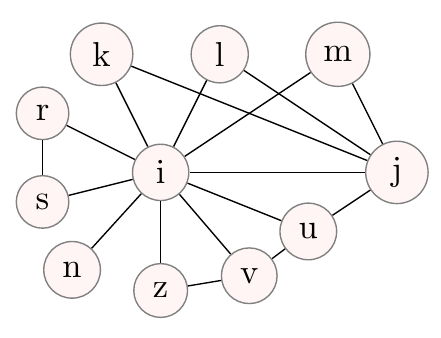}
	\caption{
		There are three different sets of 1-neighbors of $i$ in terms of gossip propagation,
		that are,
		$
		   \Gamma_{k}(i) 
		= \Gamma_{l}(i)
		= \Gamma_{m}(i)
		= \Gamma_{j}(i)
		= \Gamma_{u}(i)
		= \Gamma_{v}(i)
		= \Gamma_{z}(i)
		= \{ k, l, m, j , u, v, z \}
		$,
		$
		   \Gamma_{r}(i)
		= \Gamma_{s}(i)
		= \{ r, s \}
		$,
		and
		$
		   \Gamma_{n}(i)
		= \{ n \}
		$.
	}
	\label{fig:TriangleCascade}
\end{figure} 

The following observation of triangular cascades will enable us to do the calculation once and reuse it.
Note that selection of originator makes no difference for a given cascade.
As visualized in \reffig{fig:TriangleCascade},
if $\ell$ receives gossip initiated by $j$,
then, in return, $j$ receives gossip initiated by $\ell$.
Therefore we have $\Gamma_{j}(i) = \Gamma_{\ell}(i)$,
which implies $\sigma_{i, \ell} = \sigma_{i, j}$.
Hence we do calculation of $\sigma_{i, j}$ only once for the entire cascade,
and use it for the remaining nodes in the cascade. 
This observation drastically reduces
the number of calculations around $i$ 
if there are triangular cascades,
which is the case in networks with community structure.

Either all neighbors of $i$ are in one cascade,
or there are multiple cascades,
every edge 
that is incident to $i$, 
has to be checked for common neighbors once,
and we repeat that  for $k_{i}$ times.
Hence the complexity is 
$k_{i} \cdot \min\{ k_{i}, k_{j} \}< \kMax^{2}$.
We need to do this for every node.
Then the complexity becomes $O(\hbAbs{V})$
if we consider $\kMax$ as a constant of real networks.

\begin{figure}[thbp]
\begin{codebox}
	\Procname{$\proc{Community-Extraction}(V, p)$}
	\li \Comment Parameters set of vertices $V$
	\li \Comment and preferred function $p \colon V \to V$
	\li \Comment Uses initially empty $stack$
	\li
	\li \Comment Set all nodes in $V$ as ``unvisited''
	\li \Comment  and label with unique node ID
	\li \While $i$ in $V$
	\li \Do
		$visited(i) = \const{false}$
		\li $community(i) = i$
	\End
	\li 	
	\li \While there is $i$ in $V$ with $visited(i) = \const{false}$
	\li \Do
		   \Comment push the nodes on the path into $stack$
		\li    \While $visited(i) = \const{false}$
		\li \Do
			    $visited(i) = \const{true}$
			\li $push(i)$
			\li $i \gets p(i)$
		\End
		\li  
		\li \Comment put nodes in $stack$ to community of $i$
		\li $C = community(i)$
		\li \While $stack$ not empty
		\li \Do
			    $j = pop()$
			\li $community(j) = C$
		\End
	\End
\end{codebox}
\end{figure}

\subsubsection{Complexity of obtaining communities}

\begin{figure}[thbp]
\begin{center}
	\includegraphics[width=0.6\linewidth]
		{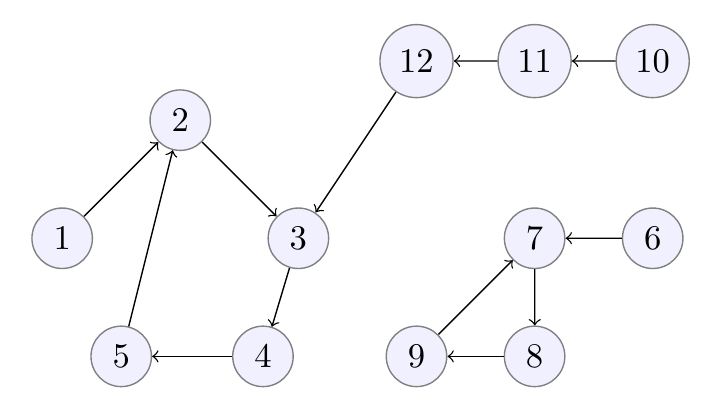}
	\caption{
		Obtaining communities from preference network.
		Initially all nodes have their IDs as labels, i.e., $C_{1}=$``1'', $C_{2}=$``2'' and so on.
		Start with node 1 and $C_{1} $= ``1'', 
		mark it as visited. 
		Follow the arcs until arriving at an already visited node.
		Arc $(5, 2)$ arrives at already visited node 2, 
		so label of community is $C_{2}$=``2'',
		backtracking resets 5, 4, 3, 2, 1 to ``2'' again in this order.
		Then get the next unvisited node in the list,
		which is node 6.
		Set it as visited and visit the nodes 7, 8, 9 until Arc $(9, 7)$ arrives at already visited node 7. 
		Community labels of 9, 8, 7, 6 are set to ``7''.
		The next unvisited node is node 10.
		Following the arcs, arc of $(12, 3)$ will arrive at already visited node 3.
		Back track labeling is triggered,
		which relabels 12, 11, and 10 as ``2''.		
	}	
	\label{fig:figCommunityExtraction}
\end{center}
\end{figure}

Now we investigate the complexity of extraction of the communities 
once the preferred function $p$ is given.
We assume that nodes have unique ID.
Initially we put all nodes in a list, mark as unvisited and 
label them with their unique ID,
i.e.
$community(i)=i$.
We process unvisited nodes in the list one by one and terminate 
when all the nodes become visited as follows.
(See 
algorithm \proc{Community-Extraction} 
and \reffig{fig:figCommunityExtraction})
We get an unvisited node $i$ from the list,
mark it as visited and push it into a stack.
Then set the preferred node of $i$ as $i$, 
i.e.,
$i \leftarrow p(i)$,
and repeat the process.
Eventually $i$ becomes a node that is already visited.
Preserve the community of it in $C$ ,
i.e.,
$C \leftarrow community(i)$.
Label all the nodes in the stack with $C$ 
while
popping out nodes from the stack.
Then repeat the process with a new unvisited node, if there is any.

Note that the algorithm passes every node twice. 
Once unvisited nodes are visited and push into the stack.
Then once more when they are popped from the stack.
So the complexity is $O(\hbAbs{V})$.

\subsubsection{Overall complexity}
Considering all the major parts of our algorithm, the overall complexity of
algorithm will include edge weight calculation and obtaining communities using preference network.
We use two methods as edge weights which require different number of calculations,
i.e.
calculating number of common neighbors used in PCN and has  $O(\hbAbs{E})$ time-complexity, 
while calculating spread capability used in PSC has  $O(\hbAbs{V})$ time-complexity.

As obtaining communities using preference network regardless 
of weight method has $O(\hbAbs{V})$ time-complexity,
the overall time-complexity of PCN algorithm is $O(\hbAbs{E} + \hbAbs{V})$
and overall time-complexity of PSC algorithm is $O(\hbAbs{V} + \hbAbs{V}) = O(\hbAbs{V})$.

\subsection{Complexity on parallel processors}

Using local information for community detection is a good candidate for parallel execution.
Let's consider the case of using common neighbors as edge weights in network.
Our algorithm has three steps of operation:
(i)~Calculate the edge weights either as the number of common neighbors or as spread capability.
(ii)~Connect the node to the node with the highest edge weight as the preferred node.
(iii)~Identify communities using the preference network.
Step (iii) is not a good candidate for parallel execution
but 
steps (i) and (ii) are perfect candidates
since each processor can do its calculations without exchanging data with another processor.

Suppose we have $P$ number of processors, 
which can run in parallel.
We can get $P$ fold speed up.
We dispatch entire network data to processors and each processor calculates one edge weight, then it calculates the preferred node for each node.

\bibliography{CDwPN}{}

\end{document}